\documentclass[superscriptaddress,aps,showpacs,twocolumn,twoside,pra,10pt]{revtex4-1}
\usepackage{amsmath} 
\usepackage{amssymb}
\usepackage{graphicx}
\usepackage{color}      
\usepackage{hyperref}   
\usepackage{natbib}
\setcitestyle{square, comma, numbers, sort&compress}

\usepackage[capitalize]{cleveref}

\def\ket#1{|#1\rangle}

\def\scal#1#2{\langle#1|#2\rangle}

\def\abs#1{\left\lvert#1\right\rvert}
\def\E#1{\cdot10^{#1}}

\def\={\!=\!}
\def\>{\!>\!}
\def\<{\!<\!}
\def\-{\!-\!}
\def\+{\!+\!}

\def\abs#1{\left|#1\right|}

\usepackage{makecell}

\newcommand{\der}{\operatorname{d\!}{}}

\newcommand{\ave}[1]{\left< #1 \right>}

\newcommand{\ii}{\mathrm{i}}

\begin{document}

\title{Quasiclassical approach to quantum quench dynamics in the presence of an~excited-state quantum phase transition}

\author{Michal Kloc}
\email[These two authors contributed equally to this work.\\E-mail addresses: ]{michal.kloc@unibas.ch, simsad@fzu.cz}

\affiliation{Department of Physics, University of Basel, Klingelbergstrasse 82, CH-4056 Basel, Switzerland}
\affiliation{Institute of Particle and Nuclear Physics, Faculty of Mathematics and Physics, Charles University, V\,Hole{\v s}ovi{\v c}k{\'a}ch 2, Prague, 18000, Czech Republic}
\author{Daniel \v{S}imsa}
\email[These two authors contributed equally to this work.\\E-mail addresses: ]{michal.kloc@unibas.ch, simsad@fzu.cz}
\affiliation{Department of Radiation and Chemical Physics, Institute of Physics, Academy of Sciences of the Czech Republic, Na Slovance 2, 182 21 Prague 8, Czech Republic}
\affiliation{Applied Physics, Department of Engineering Sciences and Mathematics, Lule\r{a} university of technology, 97187 Lule\r{a}, Sweden}
\author{Filip Han\'{a}k}
\affiliation{Department of Radiation and Chemical Physics, Institute of Physics, Academy of Sciences of the Czech Republic, Na Slovance 2, 182 21 Prague 8, Czech Republic}
\author{Petra~Ruth~Kapr\'{a}lov\'{a}-\v{Z}\v{d}\'{a}nsk\'{a}}
\affiliation{Department of Radiation and Chemical Physics, Institute of Physics, Academy of Sciences of the Czech Republic, Na Slovance 2, 182 21 Prague 8, Czech Republic}
\author{Pavel Str\'{a}nsk\'{y}}
\author{Pavel Cejnar}
\affiliation{Institute of Particle and Nuclear Physics, Faculty of Mathematics and Physics, Charles University, V\,Hole{\v s}ovi{\v c}k{\'a}ch 2, Prague, 18000, Czech Republic}

\date{\today}

\begin{abstract}
The dynamics of a quantum system following a sudden, highly non-adiabatic change of its control parameter (quantum quench) is studied with quasiclassical techniques.
Recent works have shown, using exact quantum mechanical approach, that equilibration after quantum quench exhibits specific features in the presence of excited-state quantum phase transitions.
In this paper, we demonstrate that these features can be understood from the classical evolution of the Wigner function in phase space.

\end{abstract}

\maketitle

\section{Introduction}
Non-equilibrium quantum many-body systems represent a very active field of research  in relation to topics such as quantum computation, quantum chaos or emergence of thermalization among others.
An experimentally feasible way how to bring a system out of equilibrium is a \textit{quantum quench}~\cite{Sen04,Cal06,Sil08,Col20}.
This protocol is implemented as an abrupt change of a control parameter $\lambda$ of the respective quantum Hamiltonian $H(\lambda)=H_0+\lambda V$ where $H_0$ is the free part whereas $V$ (where $[H_0,V]\neq 0$) represents interaction whose strength is controlled.
Such a protocol can be routinely engineered these days, for instance, using cold atoms in optical traps~\cite{Kau06,Hof07,Che12,Tro12,Mei13,Lan13,Lan15}.
In recent years, the progress on the experimental side has been  accompanied by a number of theoretical studies~\cite{DeGra10,Per11,Mon12,San15,San16,Tav16,Jaf16,Ber17,Jaf17,Tav17,Tor18,Klo18,Col18,Col18b,Mit18,Kas18,Pas19,Jaf19,Gei20}, addressing the above mentioned topics. 


The practical realization of the protocol is the following.
First, we prepare the system $H(\lambda_{\rm i})\equiv H_{\rm i}$ in its eigenstate $\ket{\psi_{\rm i}}$ for an \textit{initial} value of the parameter $\lambda_{\rm i}$. 
Then a rapid change of the control parameter to the new value $\lambda_{\rm f}$ is performed; hence, the initial state further evolves with a new \textit{final} Hamiltonian $H(\lambda_{\rm f})\equiv H_{\rm f}$.
After some transition period the system approaches equilibrated regime characterized by time-independent mean values of the observables~\cite{Tav17,Klo18,Nic19}.

It is known that the equilibration process can be influenced by excited-state quantum phase transitions (ESQPTs)~\cite{Per11,San15,San16,Ber17,Klo18,Cej20}.
These are generalizations of ground-state quantum phase transitions (QPTs)~\cite{Car10,Sac11} and primarily manifest as singularities in the level density of the excited states.
They appear mostly in models describing some collective features of interacting many-body systems.
In such cases, the infinite-size limit (thermodynamic limit) corresponds with the classical one $\hbar \to 0$ and ESQPTs  are  associated with the presence of stationary points in the classical version of the Hamiltonian $H \to H_{\rm cl}$~\cite{Cej06,Cap08,Cej08,Str14,Str15,Str16,Mac19}.

In Ref.~\cite{Klo18} a detailed analysis of the ESQPT-induced effects on quench dynamics was performed within the class of such models derived from the Dicke model~\cite{Dic54,Tav68,Bra13}.
It was shown that the role of ESQPTs becomes significantly important if the dynamics is regular or weakly chaotic.
In these cases the signatures of ESQPTs are clearly captured in the time evolution of survival probability of the initial state $\ket{\psi_{\rm i}}$ as well as in the time evolution of the observables.
These signatures, however, depend on the details of the quench protocol as well.

Linking the specific features in the evolution of the survival probability of $\ket{\psi_{\rm i}}$  with the critical properties of the spectrum of $H_{\rm f}$ was one of the main results in Ref.~\cite{Klo18}.
In this paper, we show that these features can be understood from the quasiclassical perspective using the classical time evolution of the Wigner function associated with the initial state $\ket{\psi_{\rm i}}$ (so-called truncated Wigner approximation), see e.g. Refs.~\cite{Hel76,Hel77,Hel81,Hil84,Ste98,Bla08,Pol10}.
The view through classical trajectories evolving in the phase space discloses the role of ESQPTs in the quench dynamics in a very intuitive way.
Moreover, we show that the quantum survival probability during the equilibration as well as a typical power-law decay can be faithfully reproduced with quasiclassical techniques.

The paper is structured as follows.
In Section~\ref{Sec:Model} we introduce the model.
The quasiclassical method and the quench protocols employed to probe the ESQPTs are discussed in the subsequent Section~\ref{Sec:QQD}.
In Section~\ref{Sec:Res} we present the numerical results.
Summary can be found in Section~\ref{Sec:Concl}.

\section{Tavis-Cummings model}
\label{Sec:Model}

\subsection{Hamiltonian and Hilbert space structure}

We consider the Tavis-Cummings Hamiltonian~\cite{Tav68} (using the convention $\hbar=1$)
\begin{equation}
H=\omega b^\dag b + \omega_0  J_z + \frac{\lambda}{\sqrt{2j}} (b^\dag J_- + b J_+)\,,
\label{Eq:TC_Ham}
\end{equation}
which is obtained from the Dicke model~\cite{Dic54} by applying the rotating wave approximation.
This model can be intuitively understood as a simplified description of interaction between quantized monochromatic light with energy $\omega$  and an ensemble of $N$ two-level atoms with transition frequency $\omega_0$. 
The bosonic  operators $b, \ b^\dag$ annihilate and create photons.
The response of the atoms to the radiation field is considered as collective, i.e., the individual atoms interact with the photons with the same phase factor.
This assumption is valid if the spatial size of the atomic ensemble is much smaller than the wavelength of the photons. 
Therefore, the system of atoms can be represented by collective quasispin operators $J_\mu=\sum_{k=1}^N \sigma^k_\mu/2$ with the symbol $\mu$ standing for $(x,y,z)$ and $\sigma^k_\mu$ representing the respective Pauli matrix acting on a $k$th atom.
The ladder operators are then constructed in a common way $J_\pm=J_x \pm \ii J_y$.
We assume full collectivity of the atomic ensemble, so the length of the quasispin $j$ is linked with the total number of atoms simply as $N=2j$~\cite{Klo17}.

The Hamiltonian~\eqref{Eq:TC_Ham} is integrable.
It means that apart from energy, there exists additional conserved quantity  which effectively reduces the number of degrees of freedom by one.
This quantity can be written as
\begin{equation}
 M=b^\dag b + J_z +j\,.
 \label{Eq:M}
 \end{equation}
It is easy to show that $[H(\lambda),M]=0$ for any $\lambda$.
It means that the Hamiltonian~\eqref{Eq:TC_Ham} conserves the total number of photons (term $b^\dag b$) and atomic excitations (terms $J_z+j$).
This symmetry forbids any interaction between the states from different $M$-conserving subspaces.
Therefore, the dynamics of the system can be studied separately in any of these subspaces.
We can also express the $M$ operator in the basis $\ket{n}\ket{m}\equiv\ket{n,m}\in{\cal H}_{b}\otimes{\cal H}_{a}$ where $n$ numbers the Fock basis in the Hilbert space ${\cal H}_b$ of photons and $m$ is the eigenvalue of the $J_z$ operator which determines the basis in the Hilbert space ${\cal H}_a$ of the atoms. 
Then we simply obtain $M=n+m+j$.
Note that we do not distinguish explicitly between $M$ as an operator and a number as it should be always clear from the context.
Considering the ranges of $n=0,1,2\ldots$ and $m=-j,-j+1 \ldots j-1,j$, we can number the individual subspaces by $M=0,1,2 \ldots$.
The dimension of these subspaces is $d=\min{(M+1,2j+1)}$~\cite{Klo17b}.

\subsection{Excited-state quantum phase transitions}
\label{SubSec:TC}
\begin{figure}[t]
	\centering
  \includegraphics[width=1\linewidth, angle=0]{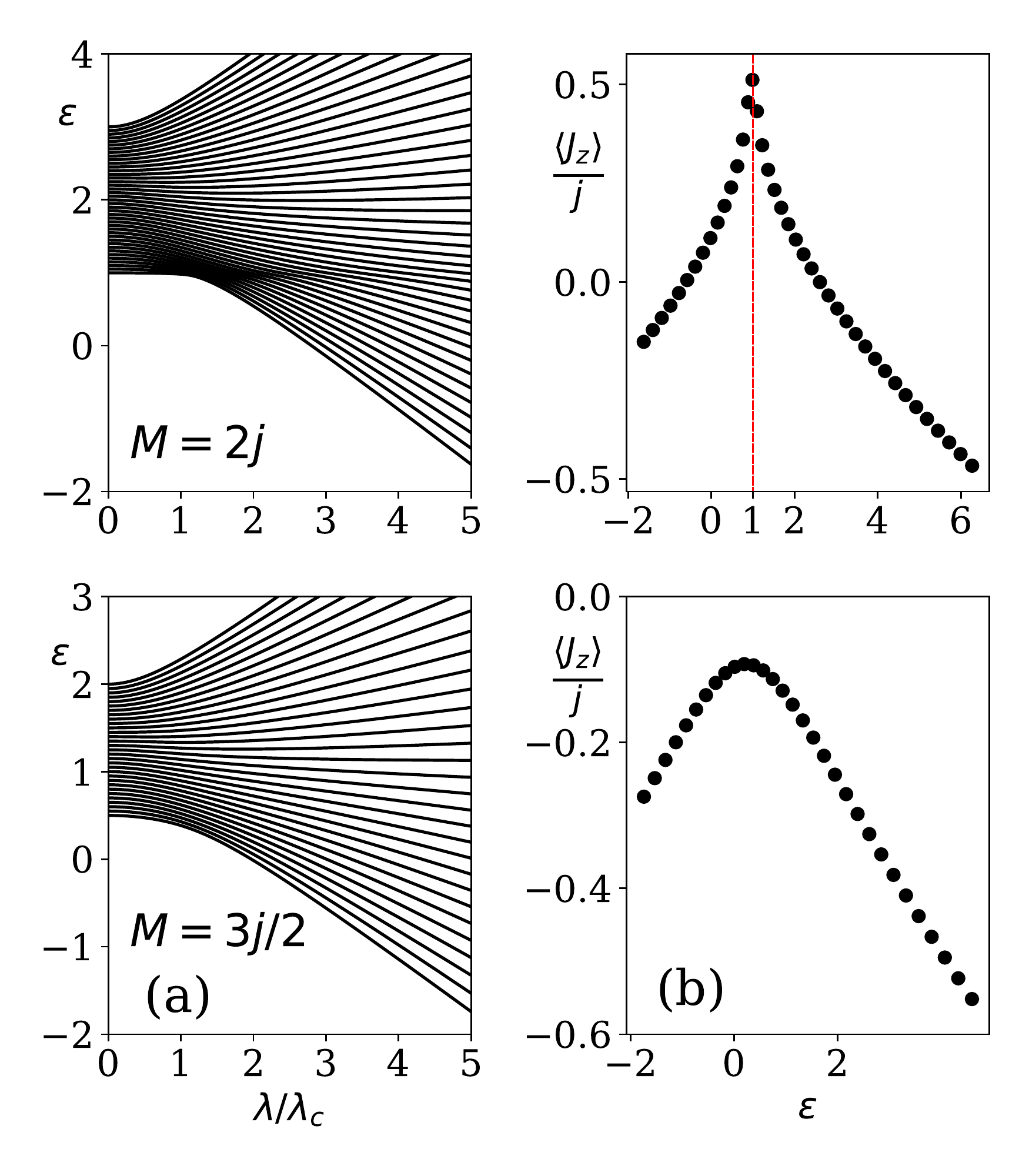}
	\caption{ Comparison of the critical subspace $M=2j$ (upper row) and a non-critical one $M=3j/2$ (lower row). Parameters used are $j=20, \omega=2\omega_0$.
	Panel (a): Energy spectra $\varepsilon\equiv E/\omega_0j$ as functions of the control parameter $\lambda$.
	Panel (b): Scaled mean value of the quasispin $z$-projection $J_z$ as a function of energy for fixed $\lambda/\lambda_c=5$ as computed from the respective spectrum in panel~(a).
	The red dashed line in the upper panel represents the the critical energy $\varepsilon_c$.}
	\label{fig1}
\end{figure}
We refer to the $M$-subspace with $M=N=2j$ as the \textit{critical} one.
The reason is that only with this specific setting of the parameter $M$ both QPT and ESQPTs appear in the spectrum~\cite{Klo17b}.
Let us define a dimensionless scaled energy $\varepsilon\equiv E/\omega_0j$.
In Fig.~\ref{fig1}(a) we show examples of energy spectra for two subspaces, the critical one (upper panel) and a non-critical one with $M=3N/4=3j/2$ (lower panel) for a moderate value of the spin length $j=20$.
The precursor of a QPT in the critical subspace is nevertheless already visible.
The dependence of the ground state energy on the control parameter $\varepsilon_{\rm g.s.}(\lambda)$ rapidly changes at the critical coupling~\cite{Klo17b}
\begin{equation}
\lambda_c=\frac{\Delta \omega}{2}\,,
\label{Eq:CritLam}
\end{equation}
where $\Delta\omega=\omega-\omega_0$ (we consider $\omega \geq \omega_0$, there is a trivial mapping between the system with this and the reversed detuning hierarchy).
On the other hand, the evolution of $\varepsilon_{\rm g.s.}(\lambda)$ is perfectly smooth in the non-critical $M=3j/2$ subspace.

For $\lambda>\lambda_c$ there is an ESQPT at energy $\varepsilon_c=1$ in the critical subspace~\cite{Klo17} which can be also anticipated directly from Fig.~\ref{fig1}(a).
Indeed, as noted earlier, ESQPTs manifest as singularities in the level density and in our case this corresponds to the sequence of avoided crossings in the vicinity of the critical point $\lambda_c$ along the energy $\varepsilon=\varepsilon_c$ for $\lambda>\lambda_c$.
However, a clear evidence is provided by Fig.~\ref{fig1}(b) where the mean value of $J_z$ in individual eigenstates is plotted against the respective eigenenergies while parameter  $\lambda/\lambda_c=5$ is fixed.
In the case of the critical subspace (upper panel), we see a sharp spike at energy $\varepsilon_c$ marking the presence of an ESQPT.
Note that similar figures like in panel~(b) with a \lq non-analytic\rq\ spike at $\varepsilon_c$ can be obtained for any $\lambda>\lambda_c$ in the critical subspace.
Here we chose $\lambda/\lambda_c=5$ because in the respective spectrum no apparent avoided crossings are visible for this coupling (which is due to the moderate value of $j$).
However, the energy dependence of $\ave{J_z}$ demonstrates the presence of the singularity anyway.

\begin{figure}[t]
	\centering
  \includegraphics[width=1\linewidth, angle=0]{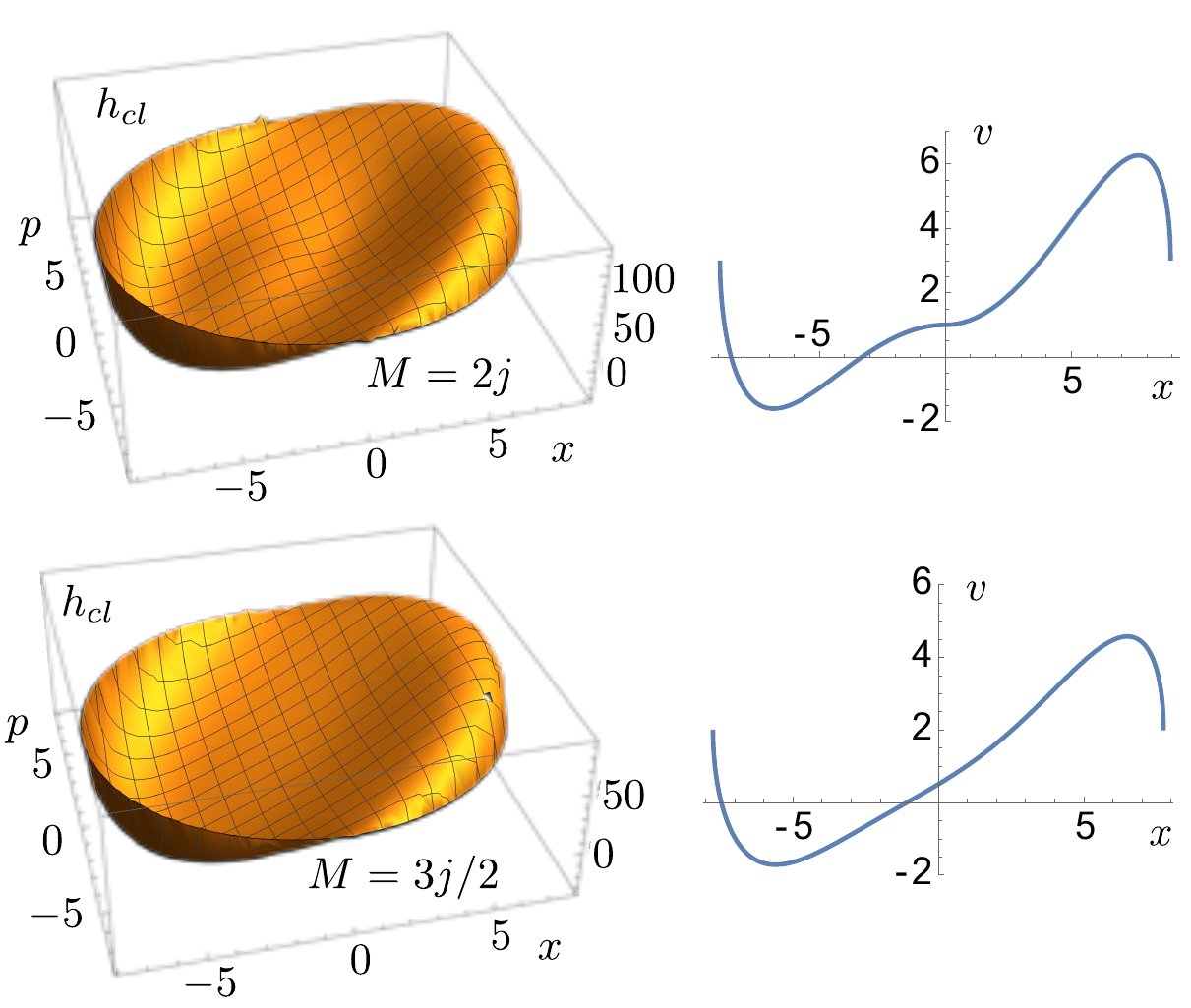}
	\caption{Comparison of the critical subspace $M=2j$ (upper row) and a non-critical one $M=3j/2$ (lower row). Parameters used are $j=20, \omega=2\omega_0$.
	Classical Hamiltonian $h_{cl}\equiv H_{\rm cl}/\omega_0 j$ in the phase space for fixed $\lambda=5 \lambda_c$ is shown together with the cut $v\equiv h_{cl}(x,p = 0)$.}
	\label{fig1b}
\end{figure}

In order to reveal the structure of the ESQPTs, let us introduce the classical version of the Hamiltonian~\eqref{Eq:TC_Ham} 
\begin{equation}
\begin{split}
H_{\rm cl}(x,p)=&\frac{\Delta\omega}{2}(x^2 \+p^2)+\omega_0 (M \-j)\\&+\frac{\lambda x}{\sqrt{j}}  \sqrt{j^2 \-\left(M \-j \-\frac{x^2 \+p^2}{2}\right)^2}\,,
\end{split}
\label{Ham_1d}
\end{equation}
where $x$ and $p$ form a  conjugate pair of classical position and momentum and $\Delta\omega=\omega-\omega_0$.
The classical limit is discussed in more detail in Appendix~\ref{Sec:appendixTC}.
An example of the (scaled) energy profile $h_{cl}\equiv H_{\rm cl}/\omega_0 j$ from Eq.~\eqref{Ham_1d} is plotted in Fig.~\ref{fig1b} for the critical (upper row) and non-critical (lower row) values of $M$.
Note that the phase space is finite.
This is due to the fact that Eq.~\eqref{Eq:M} limits the expression $x^2+p^2\leq 2 M$.
The respective one-dimensional cut at zero momentum $v(x)\equiv h_{cl}(x,p=0)$, defining a \lq quasipotential\rq , is also shown.
The stationary point at the corresponding energy $\varepsilon_c$ is directly visible in the critical subspace.

Let us comment on the phase space structure of the $M$-subspaces with $M>2j$.
Due to the square root in Eq.~\eqref{Ham_1d} a circular hole opens around the origin $[x,p]=[0,0]$.
This is also clear from Eq.~\eqref{Eq:M} if expressed in the $\ket{n,m}$ basis as $M=n+m+j$.
The photon number $n$ is linked with $x,p$ in a standard way as $n=(x^2+p^2)/2$ (see Appendix~\ref{Sec:appendixTC}).
When $M$ is strictly larger than $2j$, the value of $n$ cannot drop to zero but only to some positive integer $n_{\rm  min}$.
Thus, the circular region $x^2+p^2\le 2n_{\rm min}^2$ of the phase space becomes forbidden~\cite{Klo17b}.
The subspaces $M>2j$ show no quantum criticality similarly to those with $M<2j$.
We, however, leave them out from our study mainly for the reason that classical propagation of the Wigner function along the inner boundary would cause numerical instabilities.
Moreover, as we aim at comparing the post-quench dynamics in similar phase space structures, differing only by the existence of a stationary point, the choice of subspaces with $M\leq 2j$ is most appropriate.

\section{Quantum quench dynamics}
\label{Sec:QQD}

\subsection{Survival probability}

The survival probability, i.e., the probability of finding the initial state $\ket{\psi_{\rm i}}$ in the evolving state $\ket{\psi(t)}$ at time $t$ after the quench, is given by
\begin{equation}
\begin{split}
P_{\rm qm}(t)&=\abs{\scal{\psi_{\rm i}}{\psi(t)}}^2=\abs{\scal{\psi_{\rm i}}{{\rm e}^{-\ii H_{\rm f}t}|\psi_{\rm i}}}^2 \\
 &=\abs{\int \der{E} \ S(E)  {\rm e}^{-\ii E t}}^2\,.
\end{split}
\label{Eq:Pqm}
\end{equation}
It can be used to monitor the post-quench evolution and disclose different regimes of the equilibration process.
Equation~\eqref{Eq:Pqm} shows that $P_{\rm qm}(t)$ is related to the Fourier transform of the so-called \textit{strength function} or \textit{local density of states} $S(E)$ \cite{Vil20}.
This quantity is defined as $S(E)=\sum_k |\scal{E_{{\rm f}k}}{\psi_{\rm i}}|^2 \delta(E-E_{{\rm f}k})$, where $k$ indexes the energy eigenstates $\ket{E_{{\rm f}k}}$ of $H_{\rm f}$, and therefore describes the distribution of the initial state in the eigenbasis of the final Hamiltonian. 
Any irregularity in the evolution of the survival probability (including those due to ESQPTs) must be reflected in the corresponding strength function, and {\it vice versa}, see Ref.\,\cite{Klo18}.
The full quantum calculation of the survival probability~\eqref{Eq:Pqm} is performed exactly, using diagonalization of the initial and final Hamiltonians in the finite-dimensional Hilbert space.

\subsection{Quasiclassical method}
\label{Sec:Method}

The evolution of the system after the quench can be modelled by the quasiclassical method based on the Wigner distribution functions in the phase space~\cite{Hel76,Hel77,Hel81,Hil84,Ste98,Bla08,Pol10}.
This approach is used here as it enables us to understand and even foresee some effects, which on the quantum level would remain uncomprehended.

The Wigner distribution corresponding to the initial state reads
\begin{equation}
 W_{\rm i}(x,p)=\frac{1}{\pi}\int_{-\infty}^\infty \psi_{\rm i}^*(x+y)\psi_{\rm i}(x-y){\rm e}^{2\ii p y} \der y\,,
\label{Eq:WignerIni}
\end{equation}
where $x, \ p$ are classically conjugate position and momentum.
The initial state is expressed as a linear combination $\ket{\psi_{\rm i}}=\sum_n c_n \ket{n,m\=M\-n}$, where $c_n$ are expansion coefficients in the $\ket{n,m}$ basis of the whole oscillator-quasispin system.
The wave function in~Eq.~\eqref{Eq:WignerIni} is then obtained as $\psi_{\rm i}(x)=\sum_n c_n \psi_n(x)$  where  $\psi_n(x)$ is the $x$-representation of the Fock state $\ket{n}$ in the oscillator space 
\begin{equation}
\psi_n(x)=\scal{x}{n}=\frac{1}{\sqrt{2^n n! \sqrt{\pi}}} { \rm e}^{-x^2/2} H_n(x)\,,
\label{Eq:IniPsi}
\end{equation} 
with $H_n$ denoting the Hermite polynomial.
The justification of this procedure is given in Appendix~\ref{Sec:appendixTC}.

To simulate the post-quench evolution in the phase space, we use the truncated Wigner approximation, in which the time dependent Wigner function $W(x,p,t)$ is calculated with the aid of the Liouville equation, i.e., solely from the classical dynamics.
For technical details see Appendix~\ref{AppTech}. 
The quasiclassical counterpart of the survival probability at time $t$ after the quench is given by the overlap between the initial and the evolved Wigner functions, namely by
\begin{equation}
 P_{\rm cl}(t)= 2\pi \int  W_{\rm i}(x,p)\ W(x,p,t)  \der{x}  \der{p} \,,
\label{Eq:QuasiclasSurPor}
\end{equation} 
where the integration is done over the full phase space.


\subsection{Quench protocols}
\label{SubSec:QProtocols}
\begin{figure}[t!]
	\centering
  \includegraphics[width=0.9\linewidth, angle=0]{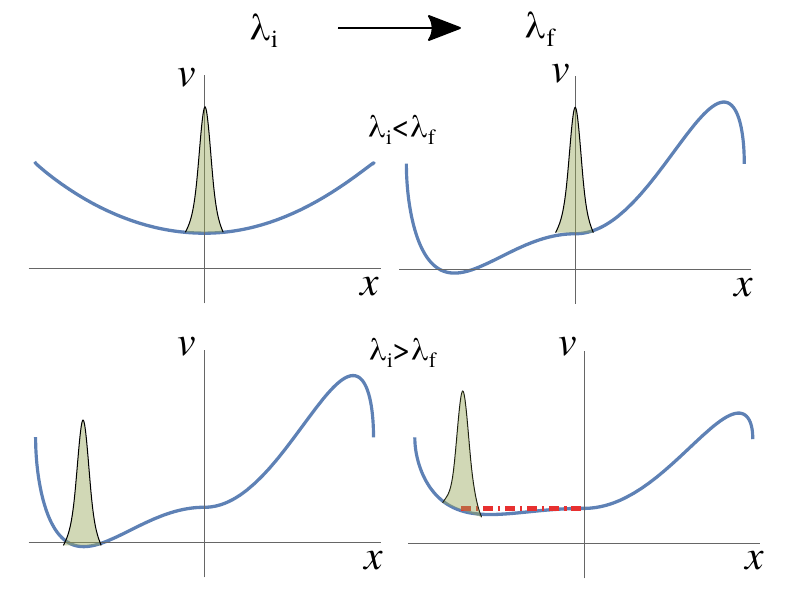}
	\caption{Sketch of the critical quench protocols used in the paper. 
	Upper row: Forward  quench critical protocol.
	 Lower row: Backward quench critical protocol. The red line indicates the equal energy of the stationary point and the value $v(x)$  at the coordinate of the center of mass of the initial Wigner function.}
	\label{fig2}
\end{figure}

\begin{figure*}[ht!]
	\centering
  \includegraphics[width=1\linewidth, angle=0]{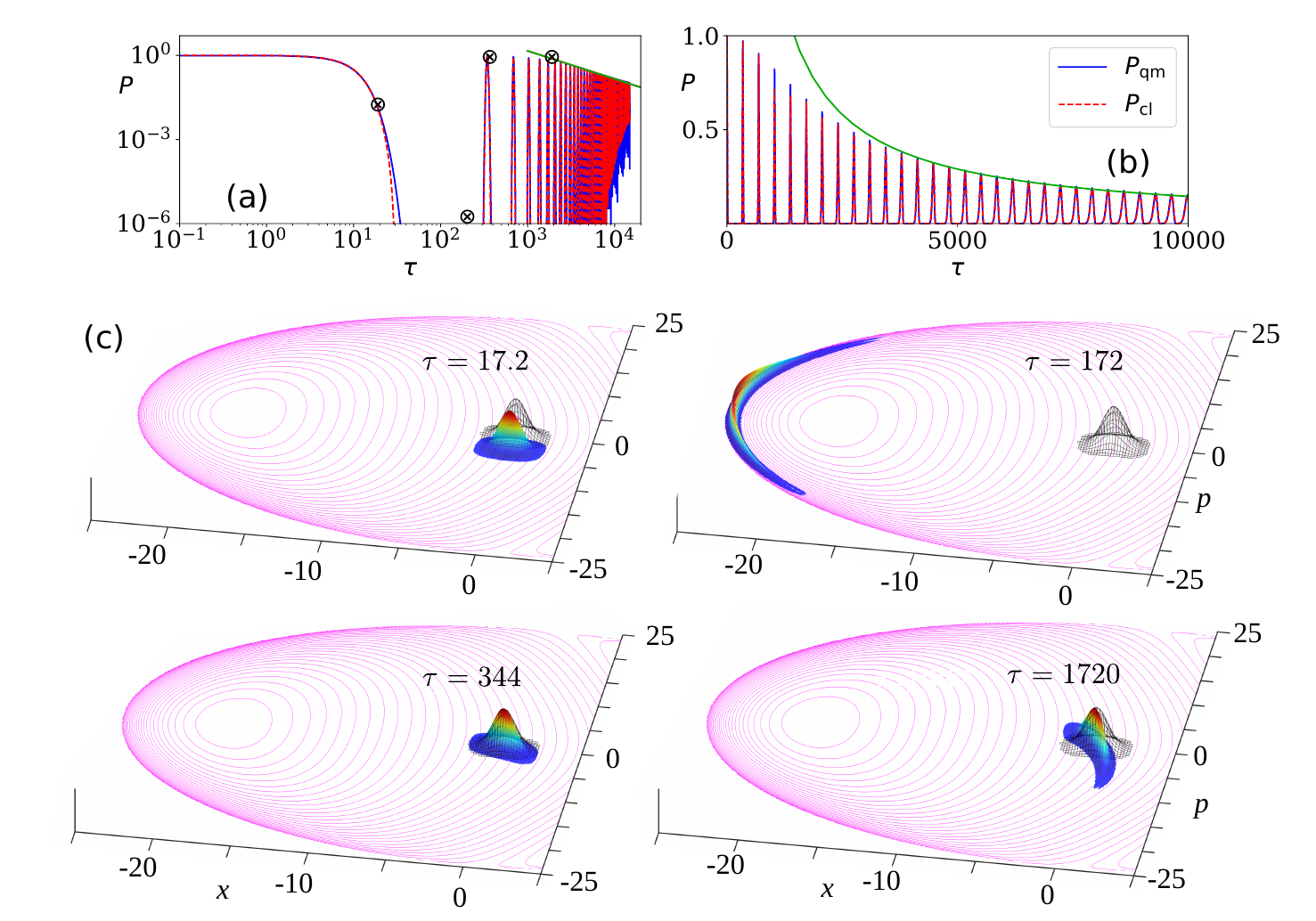}
	\caption{Forward quench in a the non-critical subspace $M=3j/2$. The parameters were chosen as $j=200, \ \omega = 2\omega_0, \ \lambda_{\rm i}=0, \ \lambda_{\rm f}=5 \lambda_c$.
	Panel~(a): Comparison of quantum and classical survival probability. The green line indicates the power-law decay $\propto 1/\tau$ of the revivals.
	Panel~(b): Same as in panel (a) with the linear scale.
	Panel~(c): Snapshots of the classically evolved Wigner function in the phase space for multiple values of $\tau$. The pink lines represent energy equivalue lines of the final classical Hamiltonian. The respective time points are marked by crosses in panel~(a).
	The initial state $\ket{\psi_{\rm i}}$ is the ground state of the initial Hamiltonian.
	The respective initial Wigner function $W_{\rm i}$ is plotted in grey.
	 Video available from Ref.~\cite{suppl}.}
	\label{fig_forwardNon}
\end{figure*}

\begin{figure*}[t!]
	\centering
  \includegraphics[width=1\linewidth, angle=0]{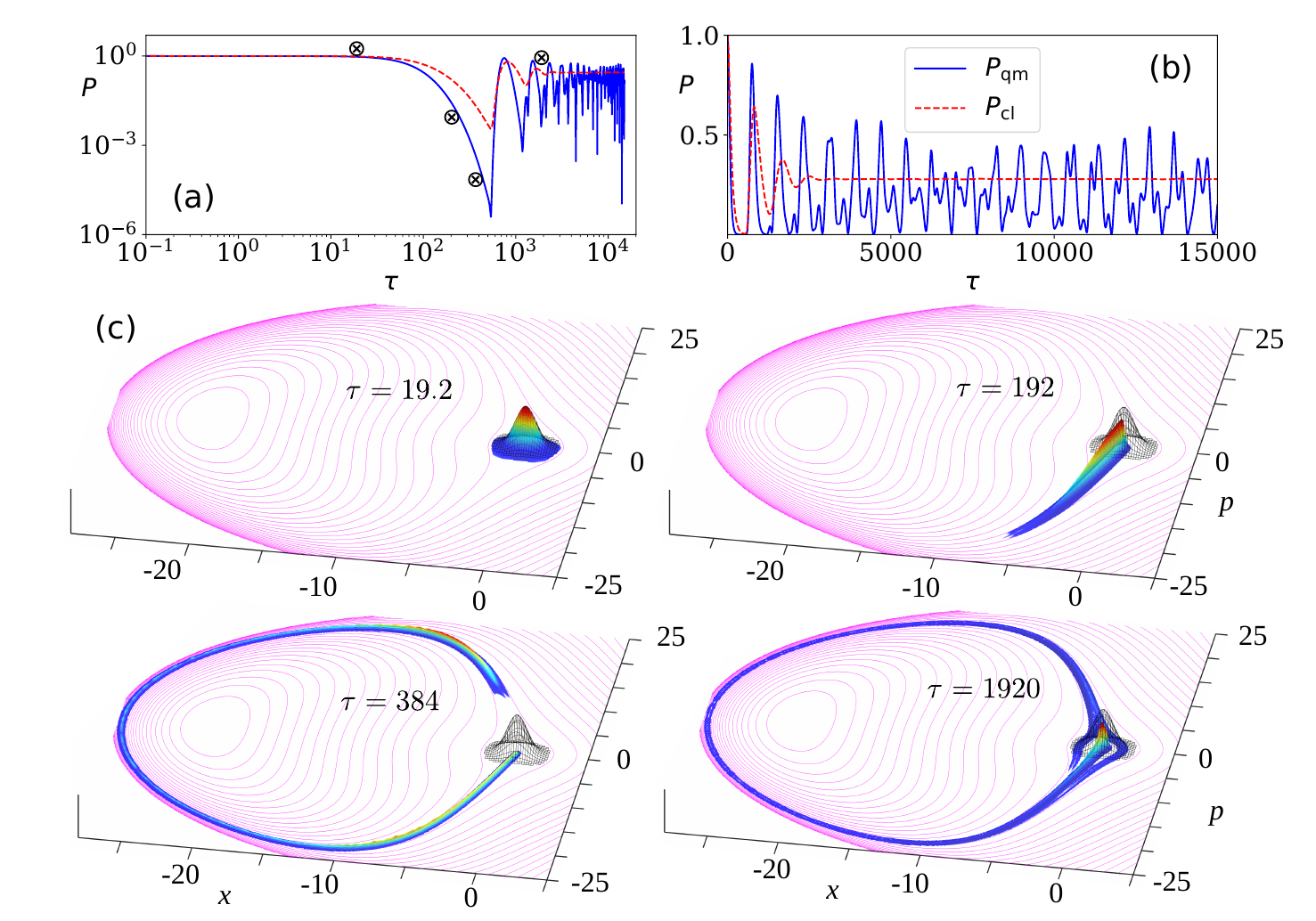}
		\caption{The same as in Fig.~\ref{fig_forwardNon}, but for the forward quench in the critical subspace $M=2j$.}
	\label{fig_forwardCrit}
\end{figure*}

In this study we always initialize the system in the ground state of $H_{\rm i}$.
We call the quench \textit{critical} if the subsequent equilibration  (i.e., time evolution according to $H_{\rm f}$) is directly affected by an ESQPT.
This happens if the support of the strength function $S(E)$ overlaps with the ESQPT critical energy $E_{\rm c}$.
In the \textit{forward quench protocols}, we start with the ground state in the critical subspace for $\lambda_{\rm i} = 0$ and then rapidly increase the interaction above its critical value $\lambda_{\rm f} > \lambda_c$.
Classically viewed, it represents a state placed initially at the bottom of the quadratic potential at $x=0$ which subsequently starts evolving due to the fact that the potential profile has been abruptly changed.
After the quench, the initial state is located at the stationary point corresponding to the ESQPT, see Fig.~\ref{fig2}, upper row.

The second option how to make the dynamics influenced by an ESQPT, is the \textit{backward quench protocol}, see the lower row of Fig.~\ref{fig2}.
We start with the system in the ground state of $H_{\rm i}$ where $\lambda_{\rm i}>\lambda_c$, so classically the state is located at the minimum energy point with $x\neq 0$.
The value of $\lambda_{\rm f}<\lambda_{\rm i}$ is chosen such that the point of the initial minimum is lifted up to the energy corresponding to the ESQPT energy (as indicated by the red dot-dashed line in Fig.~\ref{fig2}).
Unlike in the previous case, the initial state is not located directly at the stationary point after the quench.
However, for a system with only a single degree of freedom the energy-conserving dynamics  makes the trajectories  explore the whole area of the phase space with the same energy (provided that it is connected). 
Therefore, the stationary point of the classical Hamiltonian will affect the time evolution at later stages when the stationary point will capture some of the trajectories, preventing their return to the initial point.  
If $\lambda_{\rm f}$  is chosen in the way that all trajectories selected by the Wigner function do not occupy the ESQPT energy, then the quench dynamics is non-critical.

The same type of the forward and backward quench protocols can be applied in any $M$-subspace simply by considering $\lambda_{\rm i}<\lambda_{\rm f}$ and $\lambda_{\rm i}>\lambda_{\rm f}$, respectively.
Of course, the critical quench protocols can be realized only for~$M\=2j\=N$.

\section{Numerical results}
\label{Sec:Res}

\begin{figure*}[t!]
	\centering
  \includegraphics[width=1\linewidth, angle=0]{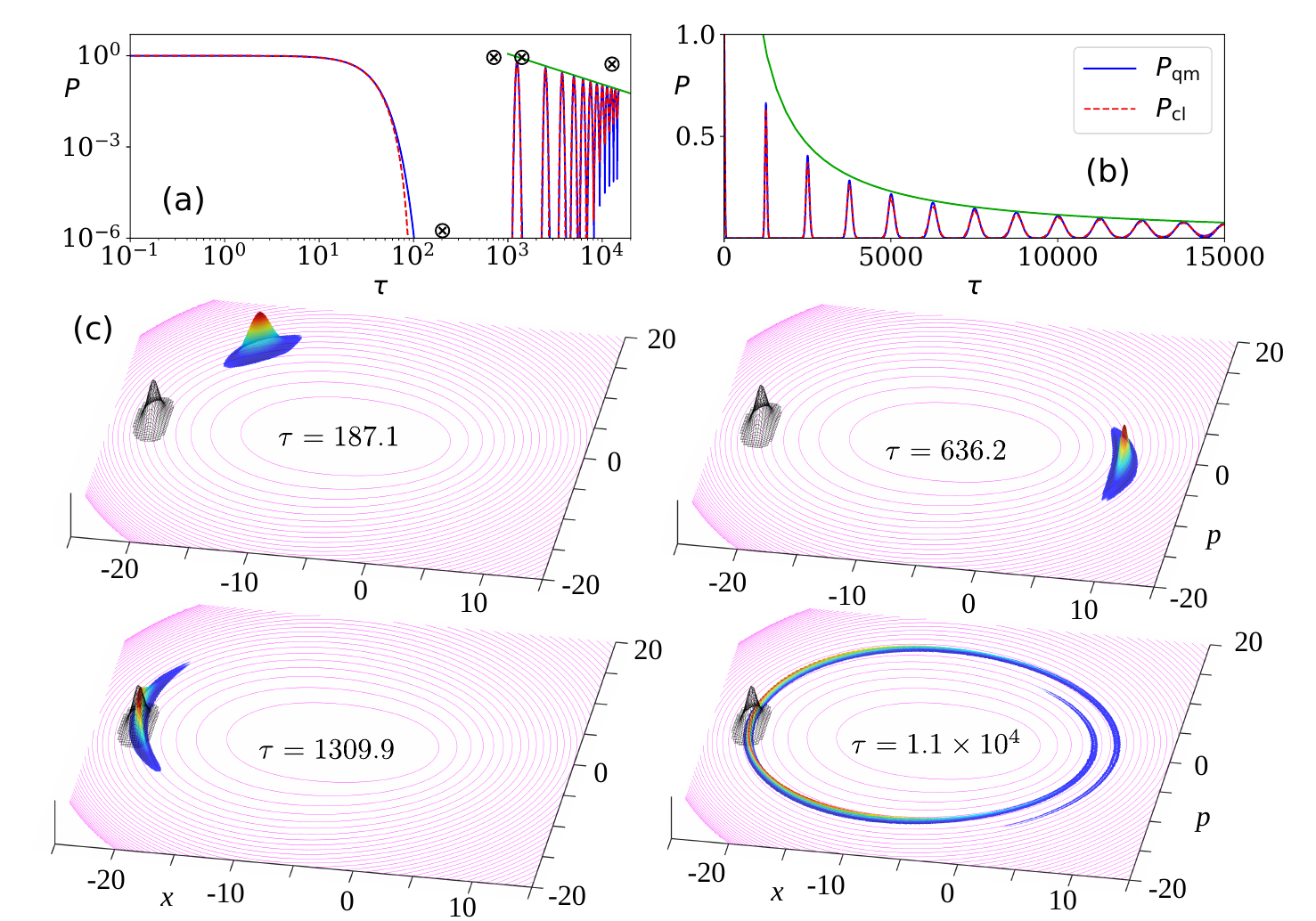}
	\caption{The same as in Fig.~\ref{fig_forwardNon}, but for a noncritical backward quench from $\lambda_{\rm i}=5 \lambda_c$ to $\lambda_{\rm f}=\lambda_c$ in the subspace $M=2j$.}
	\label{fig_backwardNon}
\end{figure*}

\begin{figure*}[t!]
	\centering
  \includegraphics[width=1\linewidth, angle=0]{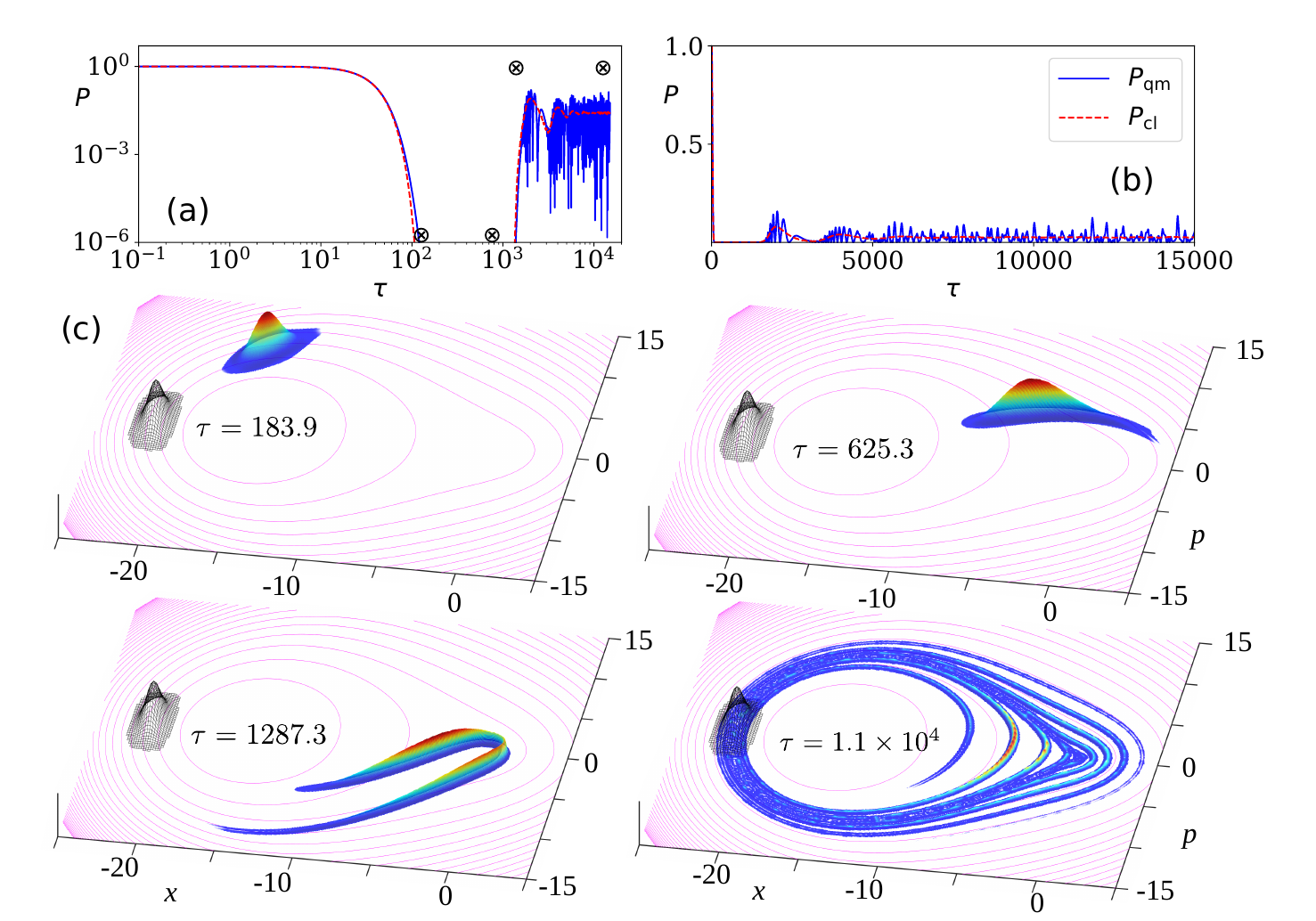}
\caption{The same as in Fig.~\ref{fig_forwardNon}, but for a critical backward quench from $\lambda_{\rm i}=5 \lambda_c$ to $\lambda_{\rm f}=1.544 \lambda_c$ in the subspace $M=2j$.}
	\label{fig_backwardCrit}
\end{figure*}

\subsection{Forward quench protocols}
\label{Subsec:Forward}

We start with the forward protocols with $\lambda_{\rm i}=0$ and $\lambda_{\rm f}=5 \lambda_c$.
From now on  we will plot all results in the time domain using a scaled dimensionless quantity $\tau \equiv (\omega_0 j)  t$.
Figure~\ref{fig_forwardNon} represents an example of a non-critical quench in the $M=3j/2=300$ subspace.
Quantum survival probability $P_{\rm qm}$ in panels (a) and (b) exhibits the standard features of \lq regular\rq\ quench dynamics of this type as described in Ref.~\cite{Klo18}.
Namely, one can identify an initial Gaussian  decay $\propto \exp{(-\alpha \tau^2)}, \ \alpha >0$, followed by sharp revivals whose amplitudes show power-law decay $\propto 1/\tau$ leading to the final equilibrated regime.

Survival probability $P_{\rm cl}$, reconstructed from the classical evolution of the Wigner function, is also plotted in Figs.~\ref{fig_forwardNon}(a) and (b) with the red dashed line.
It faithfully reproduces the quantum version. 

In Fig.~\ref{fig_forwardNon}(c) some snapshots of the classically evolved Wigner function in the phase space are presented.
Initially at $\tau=0$, the Wigner function has a Gaussian shape as it  effectively corresponds to the ground state of a simple harmonic oscillator.
It further evolves along  closed trajectories representing equivalue lines of the classical Hamiltonian.
At short time scales around $\tau\approx 17.2$, the Gaussian shape is still preserved and the classically evolved Wigner function is slightly displaced, which leads to a  gradual decrease of $P_{\rm cl}$.
For $\tau=172$, there is a zero overlap with the initial $W_{\rm i}$.
The shape is also modified as different trajectories propagate with different velocities.
The first revival appears at circa $\tau=344$ when the first period is completed.
Note that the original Gaussian shape has been almost completely recovered, which corresponds with a nearly perfect revival $P_{\rm cl} \approx 1$.

This scenario further repeats itself.
Transition to the equilibrated regime is related to the fact that individual trajectories dephase as time grows.
As a result, the recurrences start to overlap which is the moment when truncated Wigner approximation  fails.
This also qualitatively explains why amplitudes of the revivals start decaying.
For $\tau=1720$ it is clearly seen that the  classically evolved Wigner function does not recover the initial Gaussian shape when recurring through the initial point in the phase space, which is a direct consequence of the above mentioned dephasing.

Specific decay exponents related to quench dynamics in interacting many-body systems have been linked with quantum chaos, the type of inter-particle interactions or even onset of thermalization~\cite{Tav16,Tav17}.
In integrable systems, it has been shown that the power-law decay $1/\tau$ results from an interplay between the Gaussian shape of the strength function $S(E)$ and its \lq regular filling\rq\ with the discrete final energy eigenstates $\ket{E_{{\rm f}k}}$~\cite{Klo18,Ler18}.
Here, we can analytically derive the $1/\tau$ decay of the revivals from the quasiclassical dynamics, see Appendix~\ref{Sec:appendix}.
The first requirement is an approximately Gaussian shape of an initial Wigner function. 
This is analogous to the Gaussian profile of $S(E)$.
The second requirement is an approximately linear change of the recurrence times with the energy of individual trajectories (Eq.~\eqref{Eq:tauE}).
This can be viewed as an analogy to the regular sampling of $S(E)$ in the final eigenbasis since the quasiclassical periods are linked to the inverse of the level spacings.
In integrable systems, the requirement on the linear energy dependence of periods for the trajectories that fill the support of the initial Wigner distribution is rather plausible.
However, a particular situation in which this requirement is not fulfilled is when some trajectories cross a stationary point of $H_{\rm cl}(x,p)$.
This is exactly the case of critical quenches with the strength function located across an ESQPT singularity of the spectrum.




Indeed, the critical forward quench in $M=2j=400$ subspace, as depicted in Fig.~\ref{fig_forwardCrit}, shows different time evolution.
The most striking difference is the long initial survival and the absence of the large region after initial decay where the survival probability would vanish as in the non-critical quench.
This is a consequence of the ESQPT, see Ref.~\cite{Klo18}.
Comparing $P_{\rm qm}$ and $P_{\rm cl}$ in panel~(a), one can see that the quasiclassical approach overestimates the quantum survival probability significantly at early stages.
Such a discrepancy is not observed in the case of non-critical dynamics (Fig.~\ref{fig_forwardNon}(a)) which leads us to a conclusion that this artifact is not explained by the semiclassical limit approximation ($j \to \infty$) nor as a numerical error.
Rather, we anticipate a genuine quantum phenomenon, such as tunnelling, taking place near the stationary point.
The comparison with the full quantum evolution is briefly discussed in Section~\ref{Subsec:quantum}.
Nevertheless the rough features of the quench dynamics (like the time of the first revival and the prolongation of the initial decay as compared to Fig.~\ref{fig_forwardNon}) are still captured within our approach.

Figure~\ref{fig_forwardCrit}(c) provides an explanation of the  ESQPT-induced effect on the survival probability from a quasiclassical viewpoint.
This time the initial Wigner function $W_{\rm i}$ at $\tau=0$ is placed directly at the stationary point of the classical Hamiltonian (this point gives rise to the ESQPT at the quantum level).
If a classical trajectory lies at exactly the same energy as the stationary point, the passage through it is infinitely slow.
Similarly, the trajectories in the adjacent region are significantly slowed down here which effectively stabilizes the initial state at the beginning.
This also leads to a rapid dephasing of the individual trajectories already during the first period, cf. Fig.~\ref{fig_forwardNon}(c) and Fig.~\ref{fig_forwardCrit}(c) at the times $\tau=192$ and $\tau=384$.
Due to the fact that a part of the  classically evolved Wigner function is stuck at the initial point, there is a small overlap for any $\tau>0$.


\subsection{Backward quench protocols}
\label{Subsec:Back}

In the backward protocols we set $\lambda_{\rm i}=5 \lambda_c$ and the initial ground state $\ket{\psi_{\rm i}}$ is obtained from the diagonalization of the Hamiltonian $H_{\rm i}$.
The respective Wigner function in $(x,p)$ variables corresponding to the classical Hamiltonian~\eqref{Ham_1d} is then computed according to the formulas Eq.~\eqref{Eq:WignerIni} and~\eqref{Eq:IniPsi}.
See  Appendix~\ref{Sec:appendixTC} for more details.
We will fix $M=2j$ so the dynamics will take place in the subspace with the ESQPT.
By precise tuning of $\lambda_{\rm f}<\lambda_{\rm i}$, the quench dynamics can be influenced by the ESQPT (critical quench) or cannot (non-critical quench).

An example of a non-critical quench is depicted in Fig.~\ref{fig_backwardNon}.
Because the initial Wigner function $W_{\rm i}$ can  be well approximated by a Gaussian, the situation is very similar to Fig.~\ref{fig_forwardNon}.
One can observe an initial Gaussian decay and subsequently a series of revivals decaying as $1/\tau$ in panels~(a) and~(b), again, see Ref.~\cite{Klo18} for details.
Panel~(c) shows a few snapshots of the classical time evolution of the Wigner function.

If the final value of the control parameter is finely tuned to $\lambda_{\rm f}=1.544 \lambda_c$, the initial Wigner function occupies a part of the phase space with the same energy as the energy of the stationary point.
This critical backward quench is depicted in Fig.~\ref{fig_backwardCrit}.
The Gaussian profile $P \propto \exp{(-\alpha t^2)}$ of the initial decay  is still present in panels~(a) and~(b).
Indeed, the stationary point does not affect the initial evolution as it is located elsewhere in the phase space.
However, when the trajectories reach that location, some get trapped there similarly as in the forward critical quench.
This leads to a rapid dephasing and so no revivals appear in the survival probability.
Faster dephasing then essentially leads to faster transition to the equilibrated regime.
Indeed, Fig.~\ref{fig_backwardCrit}(c) for $\tau=1.1 \times 10^4$ shows that most of the available phase space is already covered with trajectories.

\subsection{Comparison between the classical and quantum evolutions} 
\label{Subsec:quantum}

\begin{figure*}[t!]
	\centering
  \includegraphics[width=1\linewidth, angle=0]{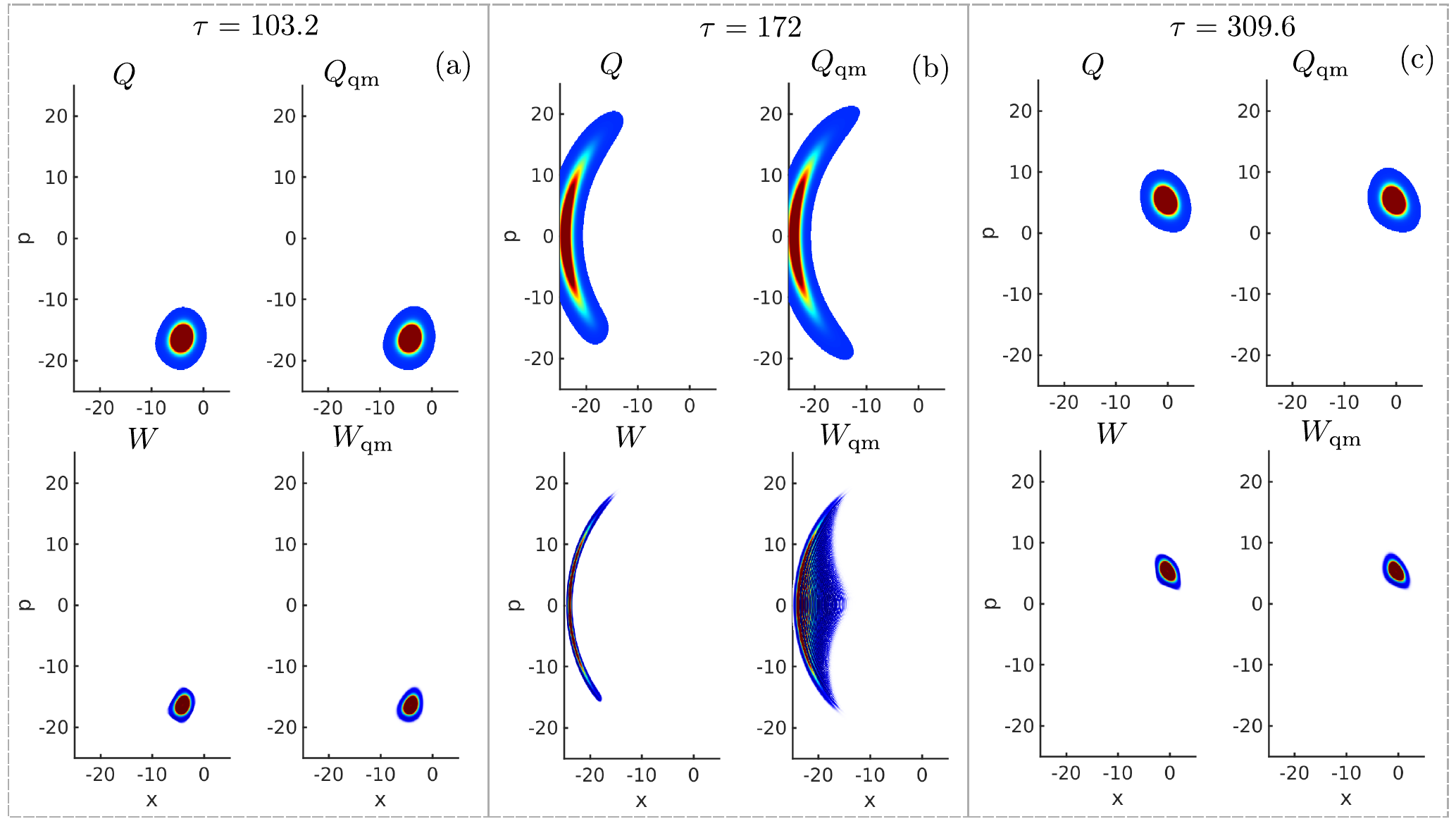}
\caption{Comparison between the classical and the full quantum evolution for the Husimi and the Wigner functions after the non-critical forward quench in the subspace $M=3j/2$.
Other parameters are the same as in Fig.~\ref{fig_forwardNon}.}
	\label{fig_quantumNonCrit}
\end{figure*}

\begin{figure*}[t!]
	\centering
  \includegraphics[width=1\linewidth, angle=0]{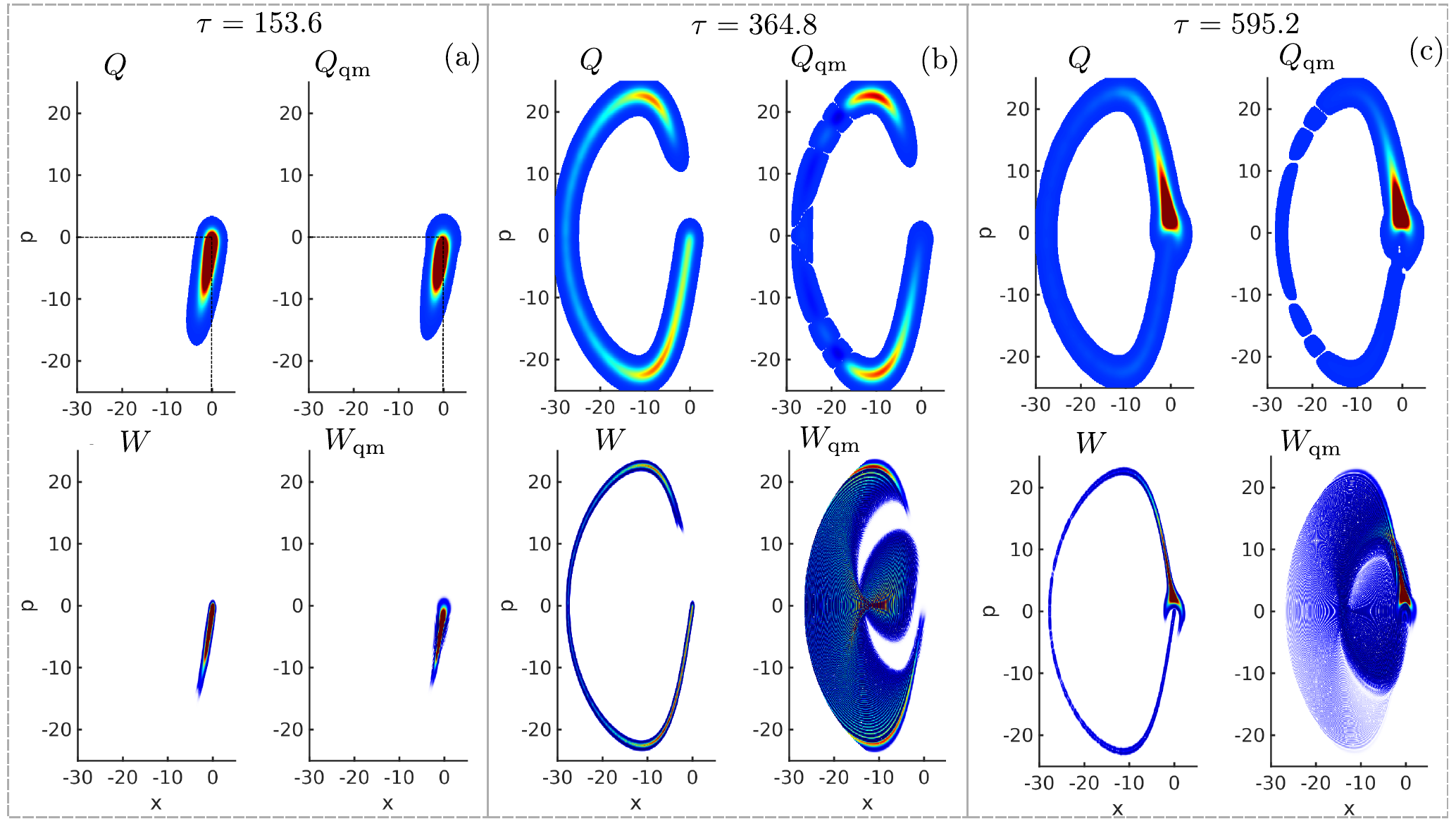}
\caption{Comparison between the classical and the full quantum evolution for the Husimi and the Wigner functions after the critical forward quench in the subspace $M=2j$.
Other parameters are the same as in Fig.~\ref{fig_forwardCrit}}
	\label{fig_quantumCrit}
\end{figure*}
We complete this section by comparing the classically evolved Wigner function $W(x,p,t)$ to its full quantum evolution $W_{\rm qm}(x,p,t)$ in the case of the forward quench.
We adopt the approach~\cite{Tak85,Tak85b} where the dynamics is compared at the level of the Husimi functions
\begin{equation}
    Q(x,p,t)=\frac{1}{\pi}\int W(x',p',t) \textrm{e}^{-(x-x')^2-(p-p')^2} \der{x'} \der{p'}\,.
    \label{Eq:Husimi}
\end{equation}
Equation~\eqref{Eq:Husimi} represents a Gaussian filter of the Wigner function over the quantum scale (recall the convention $\hbar=1$).
Thus, the quantum-evolved Husimi function $Q_{\rm qm}$, which is always positive definite and smooth, resembles more faithfully the smoothed distribution of classical trajectories.
Importantly, this rule is applicable even in the cases where oscillatory parts develop in the quantum-evolved Wigner function $W_{\rm qm}$ and the quantum Wigner distribution becomes very different from the distribution of classical trajectories.

{In Fig.~\ref{fig_quantumNonCrit} we plot the snapshots of the time evolution of the Wigner functions after the forward non-critical quench as studied in Sec.~\ref{Subsec:Forward}.
$W_{\rm qm}$ in panel (b) shows typical quantum interference pattern between the tails of the distribution.
At the level of Husimi functions the quantum and classical evolutions show a very good agreement.
This is in compliance with the results depicted in Fig.~\ref{fig_forwardNon}(a) and (b) which show that the quantum survival probability is obtained with high accuracy based on classically evolved Wigner function.

As expected, quantum effects are more pronounced in the critical forward quench, see Fig.~\ref{fig_quantumCrit}.
Due to quantum interference the wave packet evacuates the initial position in the phase space $[x,p]=[0,0]$ faster than in the case of the classical evolution.
This can be seen from panel (a) and corresponds with our previous findings from Fig.~\ref{fig_forwardCrit}(a) and (b) where $P_{\rm qm}$ decreases more rapidly than $P_{\rm cl}$ during the initial decay.
Yet another interesting effect is observed by comparing $Q$ and $Q_{\rm qm}$ for the critical quench.
The classical evolution does not allow the packet to split into two or more.
This is however possible in the quantum case due to negative interference. 
We assume that this happens in panels (b) and (c) for $Q_{\rm qm}$ as it seems to vanish (within the numerical precision) near the left turning point.
The classically evolved counterpart $Q$, as
expected, remains fully connected.
Despite the genuinely quantum effects that have impact on the dynamics of the critical quench, we can conclude that the main features are still faithfully captured by the classical evolution, e.g. the evolution of the maxima of $Q$ and $Q_{\rm qm}$.

The comparison in the case of the backward critical and non-critical quench would have many similar features.
Thus, we verify that results obtained in Secs.~\ref{Subsec:Forward} and \ref{Subsec:Back} using truncated Wigner approximation are relevant for the quantum case.

\vspace{0.2cm}
\section{Summary}
\label{Sec:Concl}

We studied quantum quench dynamics in the class of integrable Dicke-type models, showing that quasiclassical technique based on the Wigner phase-space distribution function can faithfully reproduce quantum survival probability during different stages of equilibration.
The method breaks down at the Ehrenfest time scale when the Wigner distribution spreads over a considerable fraction of the accessible phase space before the onset of the equilibrated regime. 
Up to that time, all our results are in agreement with those in Ref.~\cite{Klo18} obtained by a pure quantum analysis.
The evolution of quasiclassical survival probability in the specific stages was directly linked with both qualitative and quantitative features of the dynamics of the  classically evolved Wigner function in phase space. 
In particular, the power law $\propto\tau^{-1}$ decay of revivals in non-critical quenches was explained by an analytic quasiclassical calculation (Appendix~\ref{Sec:appendix}).

The quasiclassical method has proven to be a suitable tool for a qualitative interpretation and intuitive understanding of different patterns in quantum quench dynamics. 
In particular for integrable systems with ESQPTs the method provides essential insight into the peculiarities of the equilibration process in the presence of classical stationary points.
The fact that the trajectories with (approximately) the same energy as these stationary points get trapped in their vicinity was shown to provide a qualitative explanation of various ESQPT-related anomalies of quantum quench dynamics observed in different quench protocols.
If the stationary point responsible for the given ESQPT is located within the support of the initial Wigner distribution (as in the forward quench protocols within our model), we observe strong stabilization of the initial state.
If, on the other hand, the stationary point is located away from the support of the initial Wigner distribution (the backward quench protocols within our model), we observe a suppression of initial-state revivals and faster transition to the equilibrated regime.
We stress that these simple conclusions can be drawn only in regular systems, and particularly in systems with a single degree of freedom.

It was already proposed in Ref.~\cite{Klo18} that dynamical \lq fingerprints\rq\ such as those reported here could be used to detect ESQPTs experimentally as their direct spectroscopic measurement is in many cases impossible.
Therefore, the intuitive view of the quench dynamics  through classical trajectories as presented in this paper brings additional benefit for interpretation of  eventual experimental results. On the other hand, considering purely classical dynamics has its limitations.
Especially close to the stationary points, quantum correlations affect the dynamics already at early stages.
So, despite the good overall match, the details of the quantum survival probability cannot be captured by our method.
This is in compliance with the standard textbook knowledge that quasiclassical approach fails in the vicinity of stationary points.
Applications of more sophisticated quasiclassical techniques in relation to the ESQPTs perhaps also in non-integrable systems remain open for future studies.

\section*{Acknowledgement}
The authors thank Milan \v{S}indelka for initiating the collaboration.
Software package QuTiP~\cite{Qutip,Qutip2} was used in the numerical analysis.
Financial support from the following resources is acknowledged: 
Swiss National Science Foundation (M.K.), NCCR Quantum Science and Technology (M.K.),  
Czech Science Foundation grants 20-09998S (P.S., P.C., M.K.) and 20-21179S (D.\v{S}., P.R.K),
Charles University project UNCE/SCI/013 (P.S., P.C.),
Czech Ministry of Education, Youth and Sports grant LTT17015 (D.\v{S}., F.H., and P.R.K.).


\appendix

\section{Classical limit of the Tavis-Cummings Hamiltonian} 
\label{Sec:appendixTC}

The classical limit of the Hamiltonian~\eqref{Eq:TC_Ham} can be obtained through the mapping
\begin{eqnarray}
\hspace{-5mm}
(J_x,J_y,J_z)&=&\left(\sqrt{j^2\!-\!j_z^2}\cos\phi,\sqrt{j^2\!-\!j_z^2}\sin\phi,j_z\right)\!,
\label{map1}\\
(b,b^{\dag})&=&\tfrac{1}{\sqrt{2}}\bigl(\tilde{x}+\ii \tilde{p},\tilde{x}-\ii\tilde{p}\bigr),
\label{map2}
\end{eqnarray}
where $(\tilde{x}, \tilde{p})$ and $(\phi, j_z)$ are two pairs of mutually conjugate coordinates and momenta~\cite{Klo17b}.
Considering that in the $j\to\infty$ limit these quantities can be replaced by ordinary numbers, we obtain the classical Hamiltonian
\begin{equation}
H_\textnormal{cl}=\omega\,\frac{\tilde{x}^2\+\tilde{p}^2}{2}+\omega_0j_z +\frac{\lambda}{\sqrt{j}}\sqrt{j^2-j_z^2}(\tilde{x}\cos\phi-\tilde{p}\sin\phi)\,.
\label{Eq:Hcl_pre}
\end{equation}
The conserved quantity~\eqref{Eq:M}, which is rewritten as 
\begin{equation}
M=\frac{\tilde{x}^2\+\tilde{p}^2}{2}+j_z+j
\label{Mcl}
\,,
\end{equation}
connects the degrees of freedom of the atom and field subsystems.
Separation of a single effective degree of freedom is achieved by the canonical transformation~\cite{Klo17}
\begin{eqnarray}
\left(\begin{array}{c} x \\ p \end{array}\right)
&=&
\left(\begin{array}{cc} \cos\phi & -\sin\phi \\ \sin\phi &  \cos\phi\end{array}\right)
\left(\begin{array}{c} \tilde{x} \\ \tilde{p} \end{array}\right),
\label{canon1}
\\
\left(\begin{array}{c}
\varphi\\ {\cal M}\end{array}\right)
&=&
\left(\begin{array}{l}
\phi+M-j\\
M-j
\end{array}\right),
\label{canon2}
\end{eqnarray}
which leads to the Hamiltonian~\eqref{Ham_1d}.
Since angle $\varphi$ does not appear in the Hamiltonian, the quantity ${\cal M}$ (and of course also $M$) is an integral of motion.

As we see in Eq.~\eqref{canon1}, the angle $\phi$ from the quasispin representation~\eqref{map1} determines the relation of the new coordinate-momentum pair $(x,p)$ describing the coupled atom-field system to the old one $(\tilde{x},\tilde{p})$ characterizing the field subsystem alone.
However, it turns out that this angle is completely arbitrary.
To show this, we first note that quantum expectation values $\ave{J_x}$ and $\ave{J_y}$ cannot determine $\phi$ as they vanish in any eigenstate of the Hamiltonian with a fixed eigenvalue of parity $P=e^{{\rm i}\pi M}$. 
The discrete symmetry under this parity transformation applies to a wide class of Dicke-type Hamiltonians including the present one. 
Moreover, the Tavis-Cummings Hamiltonian~\eqref{Eq:TC_Ham} possesses also a continuous symmetry under the transformation $U(\alpha)=e^{\ii\alpha(M-j)}$ with arbitrary angle $\alpha\in[0,2\pi)$. 
This results in a gauge rotation $(b,b^\dag)\mapsto(e^{-\ii\alpha}b,e^{\ii\alpha}b^\dag)$ of the boson operators and a simultaneous counter-rotation $J_\mu\mapsto e^{{\rm i}\alpha J_z}J_\mu e^{-{\rm i}\alpha J_z}$ of the quasispin operators with $k=x,y,z$~\cite{Klo17b}. 
In fact, the conservation of quantity $M$ follows from this symmetry.
In the coordinate-momentum form the transformation reads as follows,
\begin{eqnarray}
\left(\begin{array}{c} \tilde{x}' \\ \tilde{p}' \end{array}\right)&=&
\left(\begin{array}{cc} \cos\alpha & \sin\alpha \\ -\sin\alpha & \cos\alpha \end{array}\right)
\left(\begin{array}{c} \tilde{x}\\\tilde{p}\end{array}\right)\,,
\label{calib1}\\
\left(\begin{array}{c} J_x' \\ J_y' \end{array}\right)&=&
\left(\begin{array}{cc} \cos\alpha & -\sin\alpha \\ \sin\alpha & \cos\alpha \end{array}\right)
\left(\begin{array}{c} J_x\\ J_y\end{array}\right)\,,
\label{calib2}
\end{eqnarray}
where formula \eqref{calib1} is of the same form as the canonical transformation \eqref{canon1}.
Therefore, any preselected value of angle $\phi$ can be altered to any other value $\phi'$, particularly to $\phi'=0$, by applying the gauge transformations in Eqs.\,\eqref{calib1} and \eqref{calib2} with $\alpha=\phi-\phi'$.

With this background we are ready to accept that the classical Hamiltonian~\eqref{Ham_1d} can be written with $(x,p)$ replaced by the original coordinate-momentum pair $(\tilde{x},\tilde{p})$.
A direct derivation avoiding canonical transformation in Eqs.\,\eqref{canon1} and \eqref{canon2} is possible via inserting $j_z$ calculated from the constraint \eqref{Mcl} into the classical Hamiltonian \eqref{Eq:Hcl_pre} with $\phi=0$. 
In view of the above considerations, this can be interpreted merely as a choice of a particular `gauge' in which $(x,p)$ of the coupled system coincides with $(\tilde{x},\tilde{p})$ of the field subsystem.

These explanations justify the procedure used in this paper to calculate the Wigner function \eqref{Eq:WignerIni}.
The coordinate and momentum operators in any fixed-$M$ subspace of ${\cal H}_b\otimes{\cal H}_a$ are written in analogy to those in the Fock space ${\cal H}_b$ of the field subsystem, cf.\,Eq.\,\eqref{map2}, but in the form that strictly conserves the value of $M$:
\begin{equation}
x\=\tfrac{1}{\sqrt{2}}(b^\dag L_-\+bL_+),\quad p=\tfrac{\ii}{\sqrt{2}}(b^\dag L_-\-bL_+).
\label{como}
\end{equation}
Here $L_{\pm}=[J^2\-J_z(J_z\pm 1)]^{-1/2}J_{\pm}$ are normalized ladder operators in ${\cal H}_a$ which compensate changes of the boson number $n$ induced by $b^\dag$ and $b$.
It can be shown that for very large $j$ and $|m|\ll j$ the operators in Eq.\,\eqref{como} approximately satisfy the expected commutation relation $[x,p]=\ii$. 
Under these conditions the relation $b^\dag J_-+bJ_+\approx x\sqrt{j^2-J_z^2}$ immediately transforms the interaction term of the Hamiltonian \eqref{Eq:TC_Ham} into its classical limit in Eq.\,\eqref{Ham_1d}. 
Using the known recursive relation for one-dimensional harmonic oscillator eigenstates $\psi_n(x)$ we can easily prove that the eigenvector of the position operator in Eq.\,\eqref{como} with an eigenvalue $x$ (the same symbol as for the operator) is given by $\ket{x}\=\sum_n\psi_n(x)^*\ket{n,m\=M\-n}$.
This results in the expression of the initial state wave function according to Eq.~\eqref{Eq:IniPsi} and the text formula above it.

\section{Technical details of the quasiclassical simulation}
\label{AppTech}
The propagation of the Wigner function $W_{\rm i}(x,p)$ from Eq.~\eqref{Eq:WignerIni} towards $t>0$ is realized according to the classical Liouville equation with the Hamiltonian~\eqref{Ham_1d} for a bunch of classical trajectories.
The initial positions of these trajectories $[x_l,p_l]$ are located on a regular grid indexed by $l$ and sampling a finite region of the phase space which is designated by non-zero values of the initial Wigner distribution.

Each trajectory is associated with a specific weight corresponding to the value of the initial Wigner distribution at that phase-space point (for our choices of initial states being the ground states of $H_{\rm i}$, the respective Wigner function has no negative components; therefore, its probabilistic interpretation is applicable).

The classically evolved distribution $W(x,p,t)$
is evaluated on a large regular grid of the $[x,p]$ space which spans the
whole region where the classical Hamiltonian is real defined.
This grid is different both in the number of sampling points
 and spacings $\Delta_x$, $\Delta_p$
from the grid $[x_l,p_l]$ which is used for
the initial positions of the trajectories.
In our computations we used a grid of the size $255 \times 250$ 
(spacing is $\Delta_x\backsimeq 0.023$, $\Delta_p\backsimeq0.024$ for forward quench and $\Delta_x\backsimeq 0.01$, $\Delta_p\backsimeq0.028$ for backward quench)
for the initial positions of the trajectories. The classicaly evolved function $W(x,p,t)$ was calculated on a grid of 
the size $400 \times 250$ ($\Delta_x = 0.0375$, $\Delta_p = 0.04$) for the purpose of evaluating the quasiclassical survival probability, 
given by Eq.~\eqref{Eq:QuasiclasSurPor}. The snapshots and videos of the classically evolved Wigner function were visualized on a grid of $500 \times 500$ and $600 \times 500$, respectively ($\Delta_x \backsimeq 0.06$, $\Delta_p = 0.1$).

 In order to get the magnitudes of $W(x,p,t)$ on the large regular grid, 
one needs a suitable interpolation between 
the points defined by the evolved trajectories $[x_l(t), p_l(t)]$
which form a sort of a dense floating web in the $[x,p]$ space.
To accomplish that we define $W(x,p,t)$ as a sum of narrow Gaussians with their centers 
given by the trajectories at time $t$, $[x_l(t),p_l(t)]$.
The width of the Gaussians corresponds to the spacing of the 
initial grid for trajectories, $\Delta_x = x_{l+1}-x_l$,
$\Delta_p = p_{l+1} - p_l$, while
the height of the Gaussians is equal to the initial weights of the trajectories given by $W_{\rm i}(x_l,p_l)\equiv W(x_l,p_l,t=0)$.

Common schemes for the time propagation are based on a power expansion of the position and momentum function in time where the higher derivatives are obtained from the previous time steps and the Hamilton equations of motion. 
Such expansions are non-converging in the case of the Tavis-Cummings model, and thus we apply a different approach. 
We make use of the energy conservation law which implies that each trajectory follows an equivalue line of the classical Hamiltonian in the two-dimensional phase space. 
First, the energy equivalue line is determined for each point $[x_l,p_l]$, which defines the initial condition for a trajectory in the phase space. 
The equivalue line is defined by a set of positions $[x_l^j,p_l^j]$,  which are determined using standard geometrical procedures applied for a two-dimensional surface of $H_{\rm cl}(x,p)$ defined on an equidistant grid in the $x$ and $p$ coordinates. 
Then the time elapsed between each two points, $t_l^{j+1} - t_l^{j}$, is reconstructed using the Hamiltonian equations of motion.

This quasiclassical approach to computing survival probability does not capture quantum correlations, and therefore is bound to fail whenever they become significant.
The breakdown occurs whenever the semiclassical van Vleck Gutzwiller propagator~\cite{van28,Gut67,Sun98,Sun98b} starts to include more then one root. 
In order to  obtain the long-time behavior, more sophisticticated quasiclassical methods including quantum corrections must be employed, see Refs.~\cite{Zda01,Zda04,Kap08}.
The specific features of equilibration related to ESQPTs appear dominantly at earlier stages of the time evolution.
Thus, the quasiclassical approach as described above represents a reliable tool for our study.

\section{Quasiclassical analytic evaluation of a non-critical quench} 
\label{Sec:appendix}

\begin{figure}[t!]
	\centering
  \includegraphics[width=0.4\linewidth, angle=0]{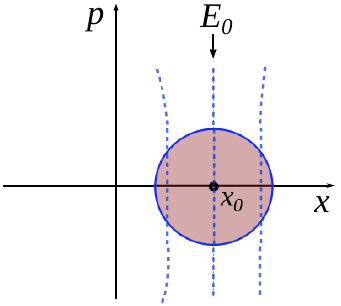}
\caption{Schematic picture of the phase space around the initial point $(x_0, p=0)$.
The circle marks the domain of $W_{\rm i}$ and the blue dashed lines represent examples of the energy equivalue lines which are  parallel with $p$ axis (approximation i.) and equidistant (approximation ii.)  in that domain.}
	\label{fig_Ap}
\end{figure}

In this appendix we derive the Gaussian initial decay as well as the $\propto 1/t$ power-law attenuation of the revivals in the case of a non-critical quench.
We assume that the initial Wigner function has a form of a Gaussian centered at point $(x_0, p_0)$ in the phase space, i.e.,
\begin{equation}
W_{\rm i}=\mathcal{N}\exp{\left[-\frac{(x-x_0)^2}{2\sigma_x^2}\right]}\exp{\left[-\frac{(p-p_0)^2}{2\sigma_p^2}\right]}\,,
\label{Eq:IniGaussian}
\end{equation}
where $\sigma_x$ and $\sigma_p$ are the respective dispersions and $\mathcal{N}$ is a proper normalization.
As the evolution always starts from rest, from now on we set $p_0=0$.
To make further calculations feasible, we employ the following approximations.
\begin{enumerate}
 \item[i.] 
 Around the initial point $(x_0,p=0)$ there is a domain of size $\sigma_x \sigma_p=\hbar$ corresponding to the quantum uncertainty.
 In the classical limit $\hbar\to 0$ the size of this region is small so we assume the energy equivalue lines to be approximately parallel with the $p$ axis within the phase space initially covered by the probability density, see Fig.~\ref{fig_Ap}.
 \item[ii.]  We also assume that the classical Hamiltonian can be approximated linearly withing this small region.
The classical potential $V(x)\equiv H_{\rm cl}(x,p=0)$ is approximately a linear function in the vicinity of $x_0$, i.e., $V(x)\big|_{x\approx x_0}=E_0 + V'(x_0)(x-x_0)$,
where $E_0=H_{\rm cl}(x_0,0)$ and $V'(x_0)=\der{V}/\der{x}\big|_{x=x_0}$.
The kinetic energy, on the other hand, is represented by a constant as $\der{H}_{\rm cl}/\der{p}=0$.
This means that in this region $H_{\rm cl}(x,p)$  is approximately constant in $p$.
\end{enumerate}

These assumptions can be justified by a detailed investigation of $H_{\rm cl}(x,p)$, from Eq.~\eqref{Ham_1d}.
As a result, we write the classical Hamiltonian around $(x_0,0)$ as
\begin{equation}
 H_{\rm cl}(x,p)=E_0+V'(x_0)(x-x_0)\,,
\label{Eq:ClasHam_Approx}
\end{equation}
where $E_0=H_{\rm cl}(x_0,0)$ and $V'(x_0)=\der{V}/\der{x}|_{x=x_0}$.

Now we define classical survival probability for a given energy shell $E$
\begin{widetext}
\begin{equation}
\begin{split}
P_{\rm cl}(t,E)&=2\pi\int W_{\rm i}(x,p)W(x,p,t) \ \delta{[H_{\rm cl}(x,p)-E]}\der{x}\der{p}\,. 
\label{Eq:ClassSurProEShell}
 \end{split} 
\end{equation}
Apparently, by integration of $P_{\rm cl}(t,E)$ over the energy domain, one recovers $P_{\rm cl}(t)$ from Eq.~\eqref{Eq:QuasiclasSurPor}.
Let $\tau(E)$ be a period of the trajectory.
$P_{\rm cl}(t,E)$ acquires non-zero values only when the trajectory is recurring through the initial region.
This means that further on we can restrict our evaluation to the times which are approximately integer multiples of the period $t \approx n\tau(E)$.
According to the assumption i., the trajectory evolves only in $p$ direction, so one can write

\begin{equation}
\begin{split}
W(x,p,t)&=\mathcal{N} \exp{\left[-\frac{(x-x_0)^2}{2\sigma_x^2} \right]}  \exp{\left[-\frac{\big(p-\dot{p}(t-n\tau)\big)^2}{2\sigma_p^2} \right]}\,, \ \ t \approx n\tau(E).
\label{Eq:Wt}
\end{split}
\end{equation}
The time change $\dot{p}$ is linked with the properties of the potential as $\dot{p}=-V'(x_0).$
We also directly inserted the time argument in the form $t-n\tau(E)$ which reflects the periodicity.

Using Eq.~\eqref{Eq:ClasHam_Approx} we define $x(E)= (E-E_0)/V'(x_0)+x_0$ and evaluate the integral
\begin{equation}
\begin{split}
P_{\rm cl}(t,E)&= 2\pi \int W_{\rm i}(x,p)W(x(E),p,t)\der{p}\\
&=2\pi \mathcal{N}^2\sqrt{\pi}\sigma_p\exp{\left[-\left(\frac{E-E_0}{V'(x_0)\sigma_x}\right)^2 \right]}\exp{\left[-\left(\frac{V'(x_0)}{2\sigma_p}\right)^2\big(t-n\tau(E)\big)^2 \right]}\,.
\end{split}  
\end{equation}

As already noted, to obtain $P_{\rm cl}(t)$ equivalent to Eq.~\eqref{Eq:QuasiclasSurPor}, we have to integrate over the energy domain.
Before we do so, we need to insert one more assumption on the function $\tau(E)$.
We will consider the simplest non-trivial dependence
\begin{equation}
\tau(E)=\tau(E_0)+\beta (E-E_0)\,,
\label{Eq:tauE}
\end{equation}
	where $\tau(E_0)$ is the period of the trajectory passing through $x_0$ (which has energy $E_0$) and $\beta$ is a real number.

Now we can write

\begin{equation}
P_{\rm cl}(t)=2\pi\mathcal{N}^2 \sigma_p \sqrt{\pi}\int_{-\infty}^\infty\exp{\left[-A^2(E-E_0)^2 \right]}\exp{\left[-B^2\big(C(t)-(E-E_0)\big)^2 \right]}\der{E}\,,
\label{Eq:Integrate}
\end{equation}
\end{widetext}
with
\begin{equation}
A=\frac{1}{V'(x_0)\sigma_x}\,, \quad B=\frac{V'(x_0)n\beta}{2\sigma_p}\,, \\ \label{Eq:AB}
\end{equation}

\begin{equation}
C(t)=\frac{t-\tau(E_0)n}{n\beta}\,.
\end{equation}
The integration \eqref{Eq:Integrate} yields
\begin{equation}
P_{\rm cl}(t)=\frac{2\pi\mathcal{N}^2 \sigma_p \pi}{\sqrt{A^2+B^2}}\exp{\left[-\frac{A^2B^2}{A^2+B^2}C^2(t) \right]}\,.
\label{Eq:ResultPcl}
\end{equation}

If we focus on the initial decay, we set $n=0$.
In that case $BC(t)=V'(x_0)t/2\sigma_p$ and $B=0$.
So the initial time evolution of survival probability is 
\begin{equation}
P^{n=0}_{\rm cl}(t) = 2\pi \mathcal{N}^2 \sigma_p \sigma_x V'(x_0) \pi \exp{\left[-\frac{V'(x_0)^2}{4 \sigma_p^2}t^2 \right]}\,, 
\label{Eq:IniDecay}
\end{equation}
which corresponds to the observed initial Gaussian decay.

Further on, let us reveal the origin of $1/t$ attenuated revivals.
The maxima  of \eqref{Eq:ResultPcl} are reached when the exponential factor becomes equal to one.
This is obtained for $t=\tau(E_0)n$ which makes coefficient $C(t)$ vanish.
For such values of $t$, the center of $W(x,p,t)$ is exactly recurring through the initial point $x_0$.
Note that now we investigate these revivals in survival probability $P^{n=t/\tau}_{\rm cl}(n)$ as a function of the number of recurrences.

Let us define the asymptotic limit $A^2 \ll B(n)^2$ which is according to Eq.~\eqref{Eq:AB} identical with the requirement given by
\begin{equation}
n|\beta| \gg
\underbrace{\frac{\sigma_p}{\sigma_x}}_{\approx 1} \frac{2}{V'(x_0)^2}\,.
\label{Eq:Ineq}
\end{equation} 
This inequality is generally fulfilled if the potential is \lq steep enough\rq\ around $x_0$, $n$ is sufficiently large, and $\beta$ has a significantly non-zero value.
Taking into account the condition~\eqref{Eq:Ineq} and the fact that $C(\tau(E_0)n) = 0$, one obtains
\begin{equation} 
P^{n=t/\tau}_{\rm cl}(n)=\frac{2\mathcal{N}^2 \sigma_p^2 \pi}{V'(x_0)|\beta|}\frac{1}{n}\,.
\label{Eq:PowerLawDec}
\end{equation} 
Recalling $n=t/\tau(E_0)$, Eq.~\eqref{Eq:PowerLawDec} shows the power-law decay $1/t$ of the revivals.
Even more, the power-law holds better for later reccurences (larger $n$), i.e., when the requirement~\eqref{Eq:Ineq} is better justified. 

In the case of the critical forward quench, this derivation fails (mainly) because the assumption ii. is not fulfilled.
Indeed, $x_0$ is a stationary point of the classical potential $V'(x_0)=0$.
In the critical backward quench, however, both the assumptions i. and ii. are reasonably fulfilled; therefore, the initial Gaussian decay~\eqref{Eq:IniDecay} is observed.
The $1/t$ power-law decay of the revivals is not observed, because the assumption on the linear scaling of the period with energy~\eqref{Eq:tauE} does not hold.
This is a direct result of the dephasing of the trajectories at the stationary point of $H_{\rm cl}$.
For analytical insight into computation of the survival probability from the quantum perspective we refer the reader to Ref.~\cite{Ler18}.
The results are in agreement with those derived in this appendix.
\\

\bibliography{RefQuasiClass_Final}

 
 %
 
\end{document}